\documentclass[useAMS,usegraphicx,usenatbib,onecolumn]{mn2e}
\usepackage{times}

\title[Stokes imaging of AM Her systems]{Stokes imaging of AM Her systems using 3D inhomogeneous models - I. Description of the code and an application to V834 Cen}
\author[Costa \& Rodrigues]
{J.~E.~R. Costa\thanks{E-mail: jercosta@das.inpe.br}
\& C.~V. Rodrigues\\
Instituto Nacional de Pesquisas Espaciais/MCT --
Av. dos Astronautas, 1758 -- 12227-010 - S\~ao Jos\'e dos
Campos - SP -- Brazil}

\begin{document}

%\date{Accepted 1988 December 15. Received 1988 December 14; in original form
%1988 October 11}

%\pagerange{\pageref{firstpage}--\pageref{lastpage}} \pubyear{2002}

\maketitle

\label{firstpage}

\begin{abstract}

Polars (or AM Her systems) are cataclysmic variables without a disc due to the strong magnetic field of the white dwarf. Most of their emission comes from the region where the accretion column impacts the white dwarf and cools through cyclotron and bremsstrahlung processes. We present a new code, CYCLOPS, to model the optical emission from these systems including the four Stokes parameters. It considers a three-dimensional region with the electronic density and temperature varying following a \textit{shock-like} profile and a dipolar magnetic field. The radiative transfer is solved in steps considering the solution with non-null input radiation. The footprint of the column in the white-dwarf surface is determined by the threading region in the equatorial plane, i.e., the region from where the flow follows the magnetic lines. The extinction caused by Thomson scattering above the emitting region is optionally included. The fittings of observational data are done using a hybrid approach: a genetic algorithm is used to seek for the regions having the best models and then an amoeba code refines the search. An example of application to multi-wavelength data of V834 Cen is presented. The fit found is consistent with previous parameters estimates and is able to reproduce the features of V834 data in three wavebands.

\end{abstract}

\begin{keywords}
novae, cataclysmic variables --  radiative transfer -- methods: numerical -- polarisation --
magnetic fields
\end{keywords}

\section{INTRODUCTION}

Cataclysmic variables (CVs) are short-period binaries consisting of a white dwarf (primary) and a late-type main-sequence star. The secondary star fills its Roche lobe, losing material to the primary by the inner Lagrangian point, $L_1$. Polars, also called AM Her systems after their prototype, are CVs in which the magnetic field reaches the 10 to 200 MG range in the primary surface. Such a strong field significantly affects the dynamics of the mass transfer preventing the formation of the accretion disc, a typical feature of non-magnetic CVs. In polars, the material from $L_1$ follows a ballistic trajectory on the orbital plane (horizontal stream) up to the magnetic coupling region, which is also called the threading region. From this region onwards, the gas-flow effectively traces the geometry of the magnetic field, forming an accretion stream which reaches the white dwarf (WD) with supersonic velocity and ionised. Another consequence of the strong magnetic field is the synchronization of the white-dwarf rotation with the orbital revolution. Reviews on polars can be found in \citet{cro90} and \citet{war95}.  

Near the white dwarf, the magnetically controlled flow creates a shocked region that emits most of the system flux in optical and X-ray wavelengths. It is usually called the post-shock region. Its location on the white-dwarf surface depends on the geometric properties of the magnetic field and the threading region distance. Considering a dipolar magnetic field, the base of the accretion stream could not be at the magnetic pole since the threading region would be too far. Specifically, given a magnetic field geometry, the distance from the threading region to the white dwarf defines the angle between the magnetic pole and the post-shock region \citep{cro89}. 
%Recently studies based on Zeeman tomography of the white dwarfs in polars suggested departures from a dipolar geometry \citep{beu07}.

The optical flux and polarisation of polars are highly modulated with the orbital phase. These observations can be explained by the cooling of the post-shock region through cyclotron process. This emission depends strongly on the viewing angle and the physical properties of the emitting region. Stokes techniques are consequently a powerful tool to diagnose the physical and geometrical properties of the accreting region. 
%The models usually assume an additional source of unpolarised and constant flux.

An important first step in the modelling of polar observations was given in '80s by different groups (\citealt{cha81,meg82,wic85}). These works assumed the emitting region as a point source with fixed values of magnetic field induction, electronic temperature, and size parameter, $\Lambda$ - this last parameter carries information about the electronic density and the optical depth of the emitting region. \citet{wic85} present a grid of fluxes, and linear and circular polarisations as function of viewing angle and harmonic for some models with different values of $\Lambda$ and electronic temperature, which can be used to fit optical data of AM Her systems. However, there are many pieces of observational evidence of an extended emitting region in polars. A natural improvement therefore is a two-dimensional modelling using a sum of point sources with distinct parameters to represent the emitting region \citep*[e.g.,][]{fer90,pot98}. This approach is based on an optically thin hypothesis in which the overlap of the emitting points do not produce moderate opacities. The density and temperature can also vary in the radial direction as a consequence of the shock. This was introduced in the models by \citet{wic88}, \citet{wuw90}, and \cite{pot04}.  Calculations on the shock structure of magnetically-channelled accretion flows can be found in \citet{cro99} and \citet{can05}.

In this work, we present a code to simulate the continuum optical emission from polars. We call it CYCLOPS, an acronym for CYCLOtron emission of PolarS. It incorporates the radiative transfer of the cyclotronic process in the post-shock region adopting a three-dimensional treatment and specific functions to describe the variation of physical quantities. For the first time, the radiative transfer is solved in a stepwise manner considering in each step the homogeneous solution of the transfer equation with non-null input radiation. The shape of the emitting region is defined by following the magnetic field lines from the threading region to the white-dwarf surface. A procedure to fit observational data is also included. In Sec. \ref{model-description}, we describe the model. Some examples of results and comparisons with previous works are shown in Sec. \ref{results}. An application to V834 Centauri's data is presented in Sec. \ref{v834}. The last section presents a summary of this work.

\section{A new model to the emission from polars}
\label{model-description}

Our objective is to model the cyclotronic emission from the post-shock region in the accreting columns of polars. It is considered as an extended three-dimensional (3D) region composed by thermal electrons immersed in a dipolar magnetic field. The code considers electron temperature and density profiles in the radial and tangential direction of the emitting region, i.e., they are not constant inside the emitting region. In this section, we describe our model as well as some details of its implementation, which has been done using as framework the {\it Interactive Data Language} (IDL).

\subsection{Defining the emitting region}
\label{sec_defining_region}

\begin{figure}
\includegraphics[width=104mm]{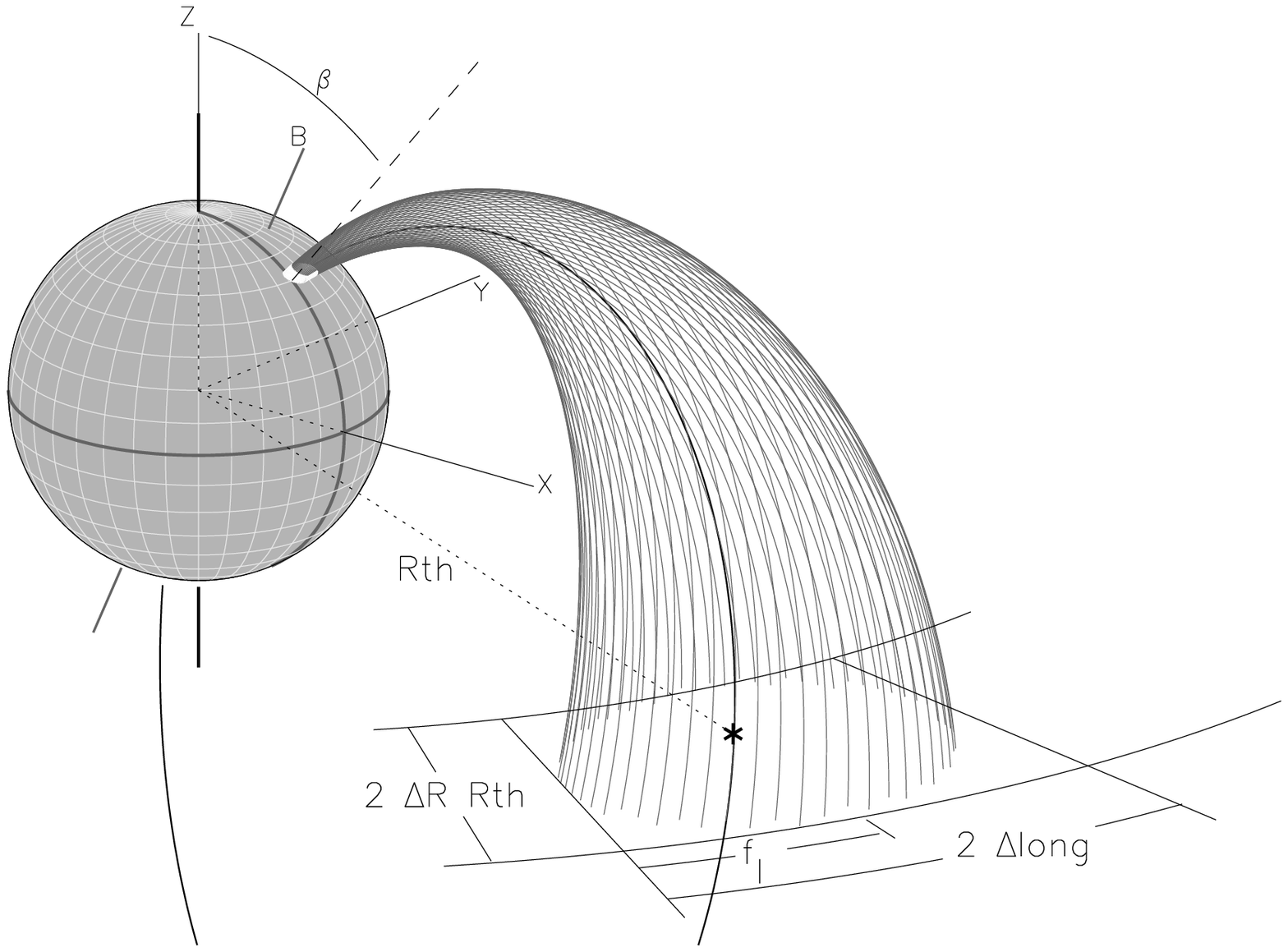} 
\includegraphics[width=64mm]{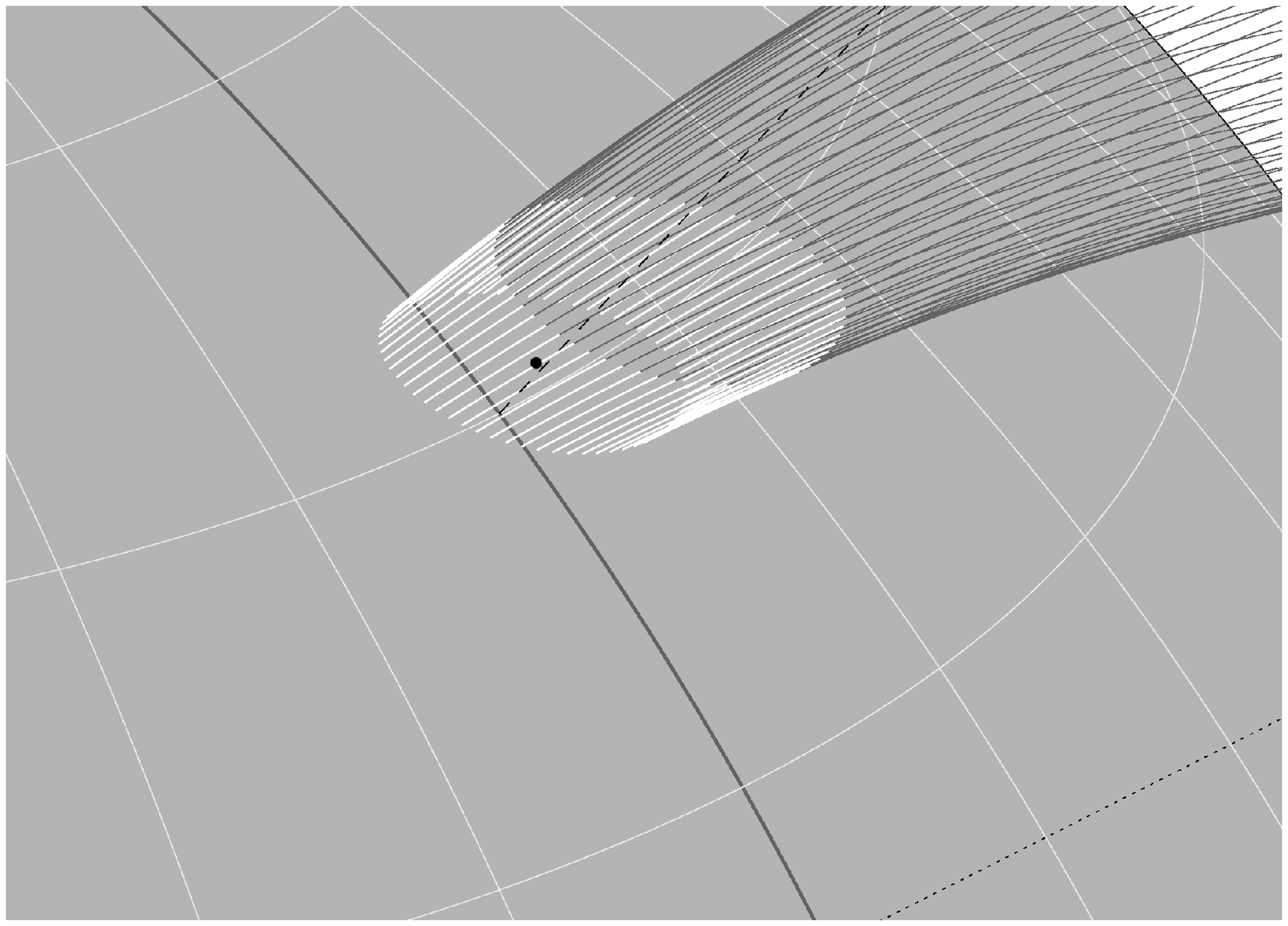} 
\caption{(Left hand panel) A diagram representing the white dwarf, the accreting column, and the emitting region (the small white region near the white-dwarf surface). It is also shown some important definitions used in the modelling. (Right hand panel) A zoom of the emitting region. The black dot represents the {\it anchor point}.}
\label{fig_diagram}
\end{figure}

In this section, we present an overview of the adopted scenario to model the optical emission coming from polars, which is schematically represented in Figure \ref{fig_diagram}. The geometry of the emitting region is defined considering a poloidal magnetic field that captures the material from the ballistic trajectory. This coupling region is called the {\it threading region}. Our model assumes that it is a 2D surface on the equatorial plane of the binary system. Below we enumerate the main definitions and assumptions of our model.

\begin{description}

\item {\sc Coordinate system.} The system as a whole is defined by a spherical coordinate system centred in the white dwarf (WD) being the polar axis parallel to the rotation axis of the white dwarf (this is also the direction perpendicular to the orbital plane). A reference point inside the emitting region defines the longitude zero. We call it the {\it anchor point}, which is represented as a black dot in Figure \ref{fig_diagram} (right). The colatitude of the {\it anchor point} is $\beta$. 

\item {\sc Inclination.} The system can be seen by any inclination, which is defined by the angle between the WD rotation axis and the observer.

\item {\sc White dwarf.} The primary is described by an opaque sphere of arbitrary radius, $R_{WD}$. The eclipse of the emitting region by the WD is consequently included. From the definition of the coordinate system, the WD angular momentum has the same direction and sense of the polar axis. The default sense rotation of the WD occurs in the counter-clockwise sense as seen by an observer at a positive latitude.

\item {\sc Magnetic field.} The direction and intensity of the magnetic field in the emitting region follow the prescription of a centred dipolar field. %An off-centre field can be easily implemented in our code, but this work presents the results to a centred dipole. 
The magnetic axis can have any angle (colatitude) to the rotation axis. Also its longitude is free. Thus, the rotation axis, magnetic axis, and accreting region are not necessarily in a same plane: this allows the magnetic field lines to depart considerably from the radial direction. 

\item {\sc Geometry of the emitting region.} %As said above, the reference point of the emitting region (the {\it anchor point}) defines the zero longitude and makes an angle $\beta$ with the rotation axis, which can assume any value between 0 and 180$\degr$. 
Its shape is of a magnetic tube whose inferior and superior ends are sections of spherical surfaces with radius $[b R_{WD}]$ and $[(b+h) R_{WD}]$, respectively. This means that a region with $ b=1$ and $h=0.1$ runs from the WD surface to 10\% $R_{WD}$ in height. The borders are defined by the magnetic field lines that delineate the threading region in the equatorial plane of the binary system as shown in Figure \ref{fig_diagram}. This procedure provides a better approach than an ad-hoc choice and a tool to, in future works, understand the threading region in polars, i.e., the region where the matter coming from the secondary is coupled to the magnetic field lines. This also prevents the choice of an emitting region resulting from an unlikely threading region. In order to calculate the "walls" of the emitting region, we first follow the magnetic field line that passes through the {\it anchor point} up to the crossing with the equatorial plane of the binary system. This defines a point so forth called the {\it threading point} at a radius of $R_{th}$ and a longitude of $long_{th}$: it is represented as an asterisk in Figure \ref{fig_diagram} (left). Then we consider a 2D threading region extending from $(1 -\Delta_R)\ R_{th}$ to $(1 + \Delta_R)\ R_{th}$ in radius and $ 2 \Delta_{long}$ in longitude. The magnetic field lines that define this boundary are followed back to the WD surface, and so define the border of the emitting region. An example of the emitting region on the white dwarf is presented in Figure \ref{fig_diagram} (right). In this procedure, $R_{th}$ and $long_{th}$ are determined by $\beta$, the colatitude of the {\it anchor point}, and the magnetic field axis latitude and longitude. The extension of the threading region accounts for the dispersion in velocities and angular momentum of the accreting matter. We would like to note that the {\it threading point} is not necessarily at the centre of threading region, which extends from $[long_{th} -2 f_l \Delta_{long}]$ to $[long_{th} + 2 (1 - f_l) \Delta_{long}]$ in longitude. In this manner, the threading point location is controlled by the parameter $f_l$ and can be at any longitude between its least and maximum values. This brings about an {\it anchor point} that can be off centre too.

\item {\sc Electronic density and temperature. } Our code allows us to adopt any analytical function to represent the matter density and temperature in the emitting region. In this work, we have considered the following options for their variation. In the tangential direction, no variation at all or an exponential decrease from the {\it anchor point} to the border of the region. As the {\it anchor point} is not necessarily centred, as explained above, the point of highest density/temperature can be displaced from the geometrical centre of the region. In the radial direction, the density decreases exponentially and the temperature increases exponentially from the WD surface to above. The used functions try to simulate the temperature and density shock solutions as shown in Figure 1 from \citet{cro99} including cyclotron cooling. However, they can be scaled in value and in height.

\item {\sc Phase.} The zero phase corresponds to the anchor point in the meridian of zero longitude. Therefore, in this phase the {\it anchor point} is located in the plane containing the white-dwarf rotation axis and the observer direction. 

\end{description}

\subsection{The radiative transfer}
\label{transfer}

In the previous section, we have described the main assumptions of our code relative to description of the emitting region. Now we turn to the details of the radiative transfer solution for the cyclotron emission. 

The radiative transfer is solved in a discrete approach as explained here. We initially define a Cartesian coordinate system in which the z axis is the observer direction, i.e., the direction of the radiation propagation. The x axis is in the orbital plane, and the y-axis completes the coordinate system. It is called the observer reference frame. A cube containing the emitting region is then defined and divided in N$^3$ voxels (elements of volume). In this representation, the emitting region is composed by (a maximum of) N$^2$ lines of sight. In each of them, the radiative transfer is solved in steps of homogeneous voxels. %(from the farthest voxel to the nearest one relative to the observer). 
Specifically, the density, temperature, and magnetic field are allowed to vary from voxel to voxel, but are constant inside a voxel. The values assumed for the calculations are that of the central point of the voxel. The number of voxels, defining the resolution of the grid, are freely chosen to properly account for the variation of the parameters and geometry of the region. The Stokes parameters that represent the integrated emission from the region are the sum of Stokes parameters of all line of sights. The voxel structure is redone for each phase in which we run the code.%, thus no interpolation is used when solving the radiative transfer.

The solution of the transfer of polarised radiation in magneto-active plasma is found in \citet[eqs. 7 to 13]{pac75} or \citet*{zhe74}. It considers a non-null incident radiation and therefore should be used 
%in a discrete approach to the radiative transfer,
here, because from the second voxel and forth the emission from the preceding voxel should be considered as input. This gives us confidence that we have an adequate treatment to any optical depth, even those intermediate from the optically thin to thick regimes. In Appendix \ref{apendice}, we present an alternative expression to the solution of \citet{pac75} and some misprints in the original formulae. In order to avoid computational problems in voxels with optical depth much smaller than 1, the formalism of \cite{pac75} for negligible absorption was used (their eq. 14). 

The \citet{pac75}'s solutions are defined in a coordinate system having one axis parallel to the component of the magnetic field perpendicular to the radiation propagation direction. We call it the plasma reference system. The magnetic field direction changes from voxel to voxel and thus the plasma reference system. Consequently this system is not appropriated to represent the emergent emission from the region as whole - see discussion presented in \citet{pac67}. The incident and emergent Stokes' vectors of each voxel are therefore represented in the coordinate system of the observer defined above. As a consequence, to solve the radiative transfer, these vectors have to be transformed to and from the plasma reference system in each voxel.

The cyclotronic emission coefficients have been calculated using the expressions from \citet{bea62} and \citet{cha80}. Our results agree very well with Figure 1 from \citet{cha80}. The absorption coefficients have been obtained from the Kirchhoff's Law, considering the thermal regime, as proposed by \citet{meg82}. The particular solution of the radiative transfer equation \citep{pac75} in this case equals to the Planck function (see Equation \ref{solution}). Following the procedure of \citet{wic85}, the bremsstrahlung absorption has also been considered. The Faraday mixing coefficients can be found in \citet{meg82} or \citet{pac77}. In this work, we have not included any scattering process in the radiative solution.

The computation is done considering a fixed frequency. Consequently, the harmonic number of the cyclotron emission varies throughout the region, because the magnetic field changes in space due to the dipolar geometry. When fitting observational data, it must be taken into account the width of the observing bandpass. We will return to this topic in Section \ref{v834}. 

In order to test the algorithm of integration by steps of the radiative transfer, we constructed regions with homogeneous physical properties (magnetic field intensity and direction, electronic density, and temperature). The resulting spectra of the stepwise solution, for different choices of parameters, was confronted to the one-step solution provided in \cite{meg82} and \cite{wic85}. The two solutions are in very good agreement.

\subsection{Extinction by the upper portion of the accreting column}
\label{sec_attenuation}

The magnetically controlled flow in polars begins at the threading region and ends at the WD surface. Our current view of polars assumes that only a portion with enhanced density near the WD, the post-shock region, emits by the cyclotronic process. There is, however, material in the flow outside the emitting region that could, in principle, attenuate this emission, if it is in between the observer and the post-shock region. This portion of the accretion column will be called the {\it attenuation region}. We have, then, included in our code the calculation of this region for which an extinction can be assumed. The attenuation region is divided in voxels and limited by the same magnetic field lines that delineate the emitting region, thus a given line of sight (from the $N^2$ ones that represent the emitting region) has above a specific number of voxels to which some attenuation can be assigned. All the voxels have the same value of attenuation, hence each line of sight has a total attenuation proportional to the number of voxels above it. The inclusion of this attenuation is optional in our code. If the user decides to use it, the program chooses the best value that fits the observed data (see next section). Each voxel has a maximum attenuation limited by the Thomson scattering cross section and the minimum density in the emitting region, which is achieved in its superior border.

\subsection{Fitting observational data}
\label{fitting}

Understanding the optical emission of AM Her systems can be divided in two main tasks: (1) to propose a model to the system and to calculate its emission; (2) to decide, among the many possible sets of model parameters, the one that best fits the observations. This section concerns the procedure to fit the data. Our code aims to reproduce the total, linear and circular polarised fluxes, and the angle of the linear polarisation as function of the orbital phase. Our code produces images of the four Stokes parameters (see Section \ref{results}). However, the observational data have not enough angular resolution to see the post-shock regions as extended objects. So the code integrates the images in one speficic phase providing a point per Stokes parameter.

A model is described by a set of parameters which are explained in detail in Section \ref{sec_defining_region}. The CYCLOPS parameters are presented as a quick list below.

\begin{itemize}
\item $i$, inclination;
\item $\beta$, the colatitude of the emitting region;
\item $b$, the distance from the WD centre to the base of the region in units if WD radius. In spite of be a free parameter in the code, we assume it equal to one in all calculations presented in this work;
\item $h$, the length of the emitting region in the WD radial direction;
\item $\Delta R$, one-half of the fractional radial extension of the threading region;
\item $\Delta long$, one-half of the azimuth extension of the threading region;
\item $f_l$, this parameter controls the position in the longitudinal direction of the {\it threading point} relative to the geometrical centre of the threading region. If the value of $f_l$ is 0 (0.5; 1.0), the {\it threading point} is at the border of smallest longitude (centre; border of largest longitude) of the threading region.  We would like to recall that this parameter controls the point of maximum density and/or temperature in the emitting region; 
\item $B$, the magnetic field intensity in the magnetic pole on the white-dwarf surface from which we calculate the magnetic moment of the dipole;
\item $B_{lat}$, the latitude of the magnetic axis;
\item $B_{long}$, the longitude of the magnetic axis;
\item $N_{e}$, the maximum value of the electronic density;
\item $T$, the maximum value of the electronic temperature.

\end{itemize}

The user must specify the frequency(ies) in which the data have been collected and is(are) not a free parameter.

There are parameters that are not part of the physical description of the system, but should be specified in the calculation. They are the number of orbital phases and the number of voxels in one dimension in the cube inscribing the emitting region (see Section \ref{sec_defining_region}). 

When fitting an observational data set, there are five quantities that are calculated for each model. We discuss each one in the following paragraphs.

\begin{description}
\item {\sc scale factor to the cyclotron flux.} The code firstly calculates the scale factor that multiplies the circular polarised flux of the model in order to provide the best match with the observed data. This flux component can only be explained by cyclotron emission and is, therefore, the best choice to fix the scale factor of the model. The same factor is then applied to the total and the linear polarised flux of the cyclotron model.

\item {\sc the contribution of an additional unpolarised source of light.} After fixing the scale of the model flux using the circular polarisation data, the code calculates the flux of an unpolarised and constant source of light that should be added to the total flux of the model to best fit the observed flux. The addition of such a source of light is usually necessary when fitting data and can be easily justified by the contribution of other components of the system, as the white dwarf, for instance. 
%This quantity varies from a set of model parameters to another.

\item{\sc an offset in phase.} The code also calculates the offset between the observed and model phases that best reproduces the data. 

\item{\sc an offset in polarisation angle.} When fitting the linear polarization, we included an offset in polarisation angle to obtain the best agreement between model and data.

\item{\sc the maximum extinction in the attenuation region.} As explained in Section \ref{sec_attenuation}, our code can include the extinction caused by the upper portion of accreting column in front of the emission from the post-shock region. If this option is set, the attenuation is fitted with the constraint of the maximum value it can assume for Thomson scattering in the {\it attenuation region}.

\end{description}

To search for the post-shock region properties in the domains of such a large number of input parameters, we have adopted a two-step approach. Initially, we use the {\it pikaia} implementation of a genetic algorithm by \citet{cha95}\footnote{http://www.hao.ucar.edu/Public/models/pikaia/pikaia.html} to look for the region in the parameter space associated with the maximum of a figure of merit related to the goodness of the fitting (g.o.f.). We then use an {\it amoeba} code \citep{pre92} to refine the search for the best model. The g.o.f. is measured by the $\chi^2$ or a modified expressions similar to it. We have implemented to weight or not each parcel of the $\chi^2$ with the observational error. The g.o.f. of a model is the sum of the $\chi^2$ for each observed curve: the total flux, the linear polarised flux, the circular polarised flux, and the position angle of the linear polarisation. For two or more bandpass data, the g.o.f. is obtained by the sum of the $\chi^2$ for each band.

\section{General results}
\label{results}

An important aspect of the emission from polars is the high variability with phase. The reason is two-fold. Firstly, the cyclotron radiative coefficient is strongly dependent on the viewing angle of the magnetic field. Second, the physical depth of the emitting region changes with phase in a region not spherically symmetric. A specific phase has, therefore, a total emission different from another phase as a result from matter distribution and view angle of the magnetic field. Both aspects of the cyclotron emission in polars is properly treated by our model. In this section, we show some results that illustrate the above discussion. 

CYCLOPS has as parameters the values of electronic density and temperature. Previous models tend to use the size parameter,  $\Lambda$. Its definition is \citep[e.g.,][]{meg82}:

$$
\Lambda = 2.01\ 10^5 \left( \frac{s}{10^5 cm} \right) \left( \frac{N_e}{10^{16} cm^{-3}} \right) \left( \frac{3\ 10^7 G}{B} \right).
$$
 
\noindent As we are using a 3D treatment, a value of $\Lambda$ does not specify a unique model: we have a family of models with different values of electronic density and height of the emitting region, $h$, that can produce a same value of $\Lambda$, supposing the magnetic field as fixed. 

\begin{figure}
\includegraphics[width=84mm,origin=c]{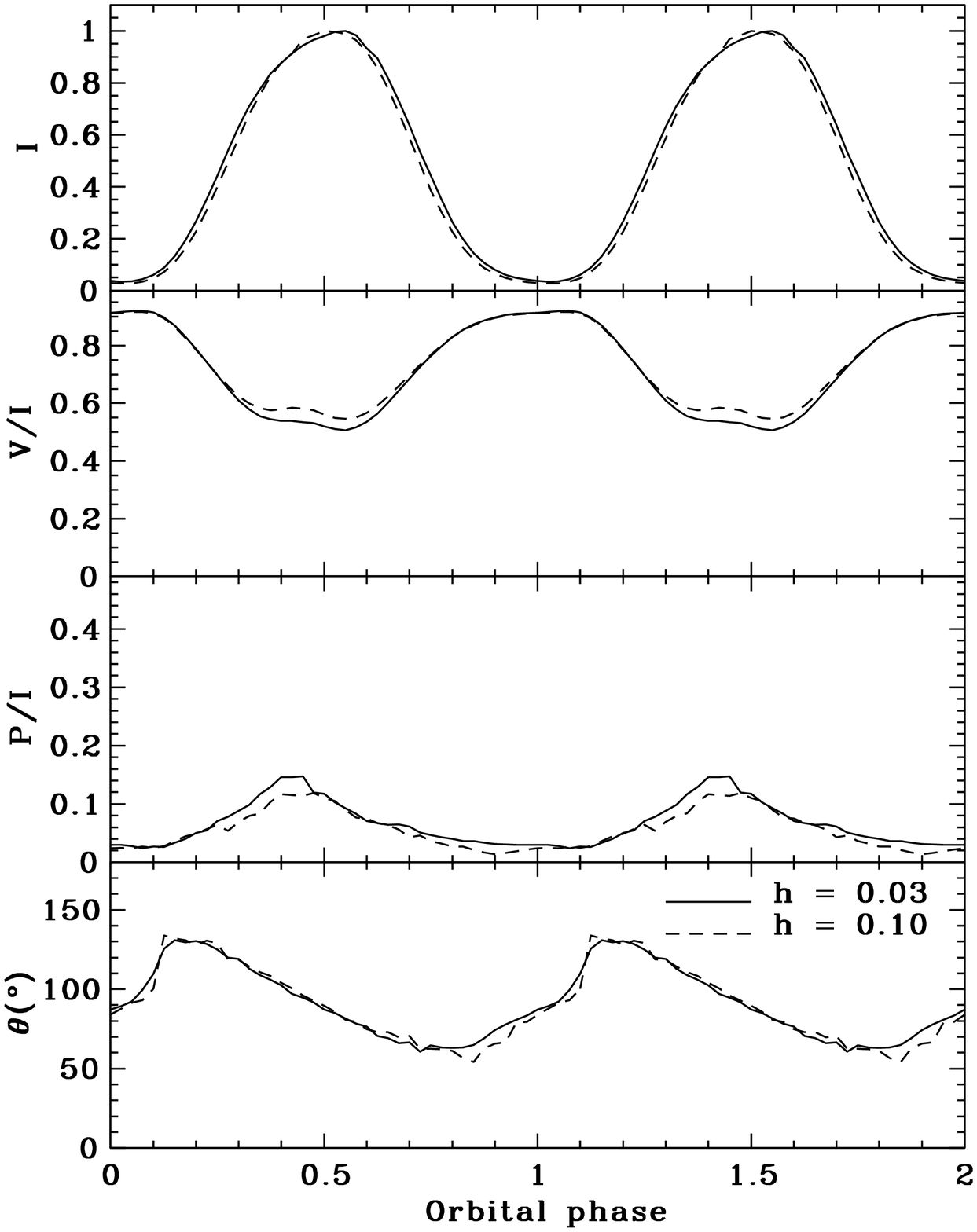}
\includegraphics[width=64mm,angle=-90,origin=c]{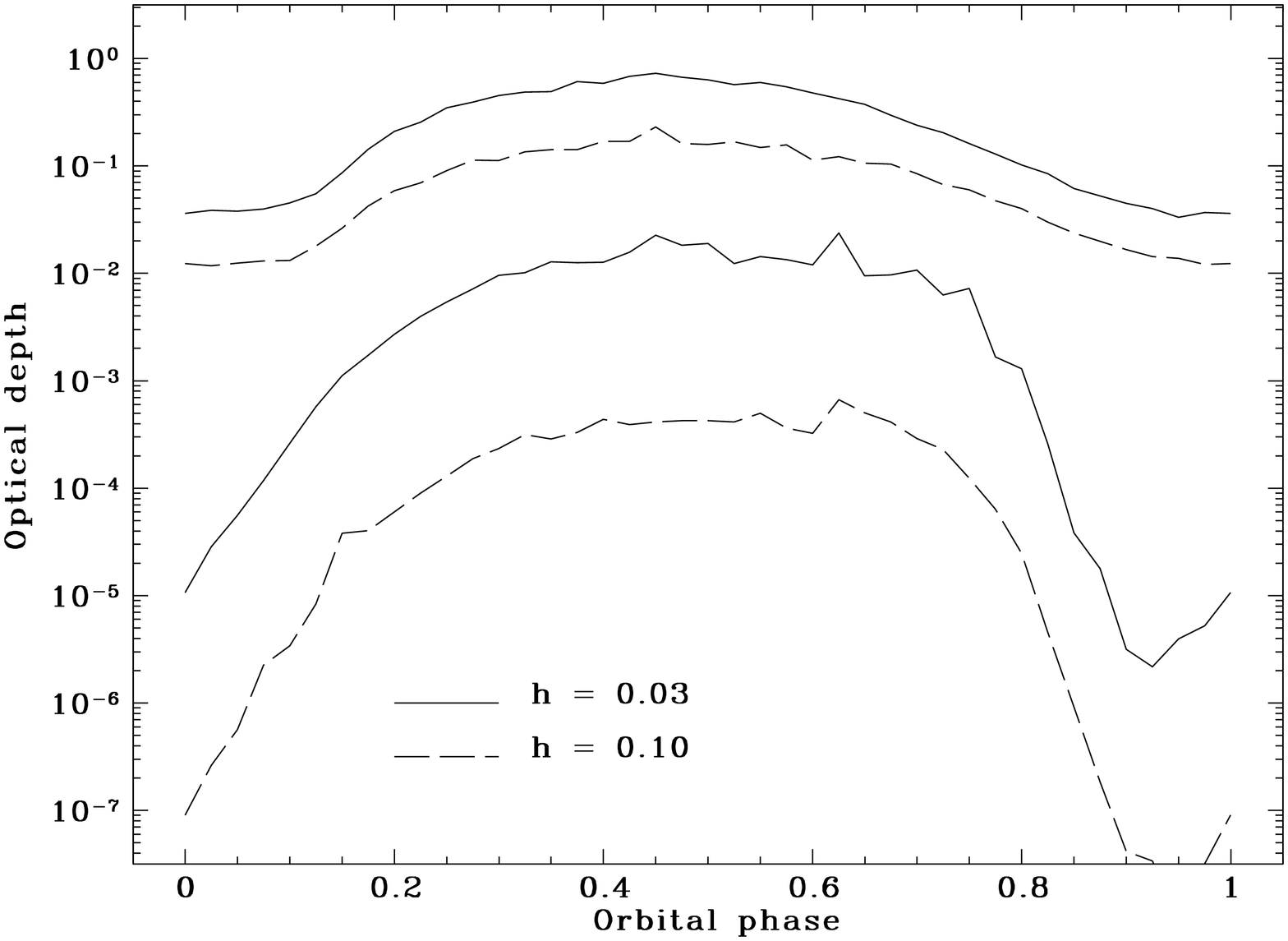}
\caption{Examples of Stokes curves for an emitting region with constant electronic temperature and density. The parameters' model are: $i$ = 50\degr; $\beta$ = 23\degr; $B$ = 29.5 MG; $T$ = 10 keV; $B_{lat}$ = 70\degr; $B_{long}$ = 29\degr; $b$ = 1.0; $\Delta R$ = 0.20; $\Delta_{long}$ = 90 \degr. (Left) The solid (dashed) line was obtained using $h$ = 0.03 (0,10) and $N_e$ = $5.13\ 10^{13}$ ($1.55\ 10^{13}$) ${\rm cm^{-3}}$. This model is equivalent to that of \citet{fer90}, figures 2 to 5, c. We have used a frequency of $6.6\ 10^{14}$ ${\rm Hz}$. (Right) The maximum and minimum values of the optical depth among the line-of-sights that compose a image as a function of the orbital phase. }
\label{fig_fw90_2c}
\end{figure}

We would like to start the presentation of our results with their validation in comparison with those of \citet{fer90}, who presented 2D extended models of polars. Figure \ref{fig_fw90_2c} (left) presents our model calculated with the parameters equivalent to the Figs. 2-5(c) from \citet{fer90}. We have adopted constant values of electronic density and temperature without the addition of a component of constant and unpolarised flux. The curves represent the results for two values of $h$, 0.03 and 0.10 ($N_e$ equals to $5.13\ 10^{13}$ and $1.55\ 10^{13}$ ${\rm cm^{-3}}$, respectively). These numbers provide the same $\Lambda$ of the models of \citet{fer90} for a radial line of sight passing in the centre of the emitting region. We have adopted a frequency of $6.6\ 10^{14}$ ${\rm Hz}$ (optical wavelengths) in order to have the harmonic number near 8. Our curves are almost identical those from \citet{fer90}. The small difference are acceptable, giving the different methodologies. The maximum optical depth (considering all the phases and all the line of sights) occurs around phase 0.5 and is 0.73 (0.23) for the model using $h$ = 0.03 (0.10) and is exactly where our two models present distinct values (Figure \ref{fig_fw90_2c}, right). That shows that the differences quickly appear as the opacity enlarges, as expected, and the optical depth depends on phase.

\begin{figure}
\includegraphics[width=84mm]{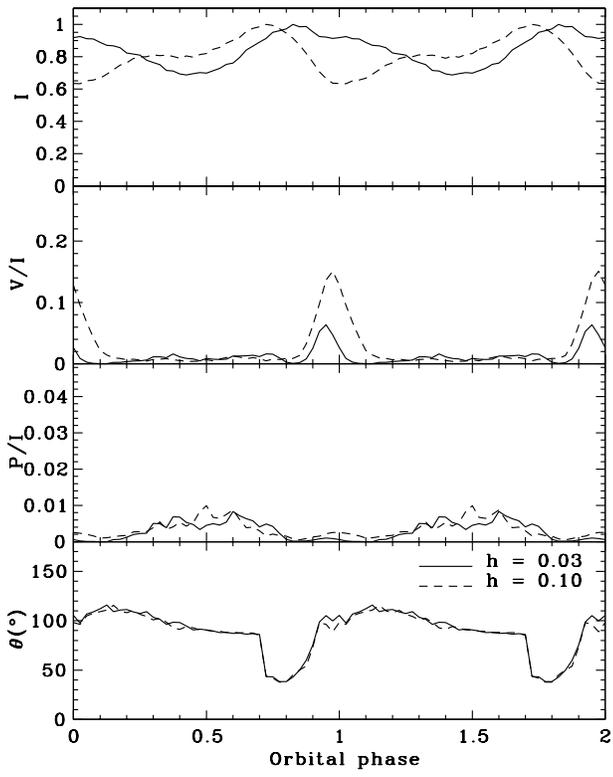}
\caption{The same of Figure \ref{fig_fw90_2c} but using a frequency of $2.0\ 10^{14}$ ${\rm Hz}$, in the near-infrared region.}
\label{fig_nir}
\end{figure}

\begin{figure}
\includegraphics[width=160mm]{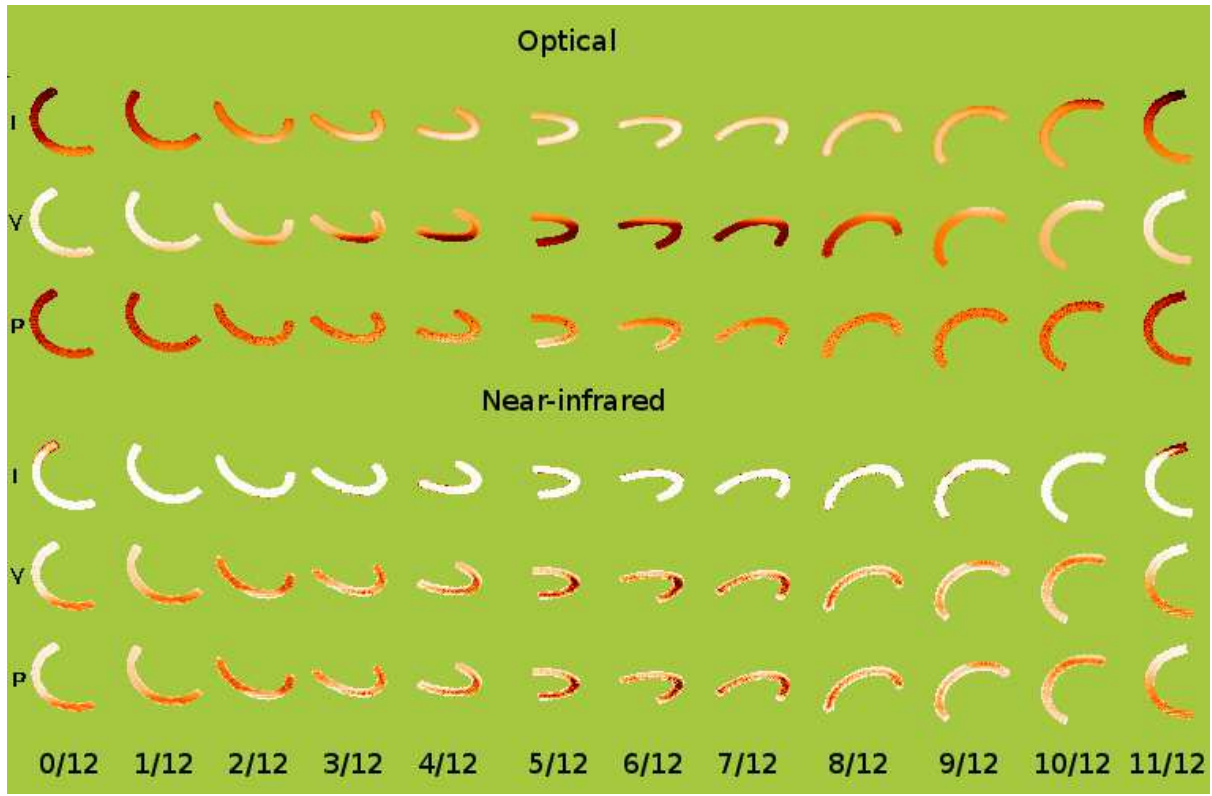}
$ $\vspace{1cm}
\caption{Stokes imaging of the models presented in Figures \ref{fig_fw90_2c} and \ref{fig_nir} for $h$ = 0.03. The three lines correspond to the flux, circular polarisation, and linear polarisation. The numbers correspond to the system phase. Each line has a differente lookup table where white stands for the highest value.}
\label{fig_image}
\end{figure}

On the other hand, the differences between our approach and 2D calculations is illustrated in Figure \ref{fig_nir}. It presents the results for the same set of parameters used in Figure \ref{fig_fw90_2c}, but for a frequency $2.0\ 10^{14}$ ${\rm Hz}$, in the near-infrared. The model samples lower harmonics and hence optically thicker region. Figure \ref{fig_nir} shows how diverse can be the curves for different values of $h$ in this regime of optical depths. Therefore a set of parameters (as those used in Figure \ref{fig_fw90_2c}) can produce similar curves in optical wavelengths, but not in the infrared. We then consider that a 3D approach is more appropriate to model the physical and geometrical properties of the post-shock regions of a polar, particularly for mildly optically thick emission. This situation is found in low harmonic numbers or in regions with high densities ($N_e > 10^{15}$ ${\rm cm^{-3}}$). Figure \ref{fig_image} shows the Stokes imaging of the previous models ($h$ = 0.03) in optical and near-infrared frequencies. It illustrates how the total flux and polarisation vary along the emitting region. For instance, at phase 0 (0/12) of the lower harmonic, the brighter region (optically thicker) is less polarized than the darker region. 

\begin{figure}
\includegraphics[width=84mm]{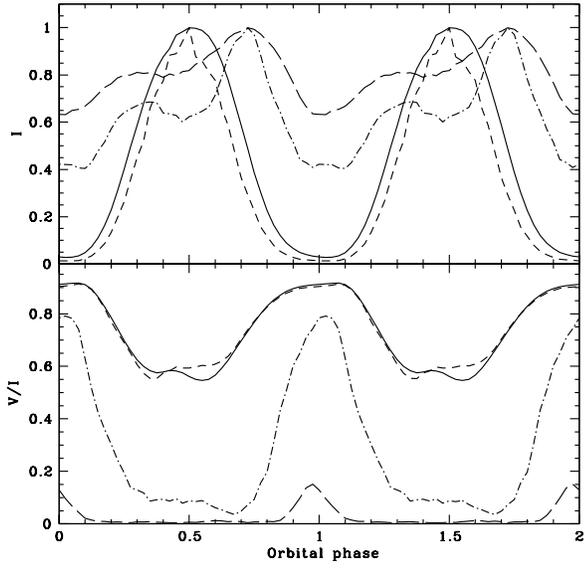}
$ $\vspace{3cm}
\caption{Comparison between models with constant electronic density and temperature and those using ``shock-like'' profiles. The constant models are represented by the solid and long-dashed lines using frequencies of $6.6\ 10^{14}$ and $2.0\ 10^{14}$ ${\rm Hz}$, respectively. The shock models are the short-dashed and the dot-dash lines for the same frequencies. The parameters are those used in the previous figures for $h = 0.10$.}
\label{fig_shock}
\end{figure}

As shock profiles are denser near the white-dwarf surface, it is interesting to check if the above result is also valid for a radially variable density and temperature. Figure \ref{fig_shock} shows our results for flux and circular polarization in optical and near-infrared wavelengths for models having (1) constant or (2) exponentially decreasing (increasing) densities (temperatures). There are differences in the curves which are striking in the near-infrared. It is noticeable the raise of the asymmetry in the flux which allow to distinguish the ``shock'' model from a shorter region (compare Figures \ref{fig_shock} and \ref{fig_nir}). Also the circular polarization is much larger than when the matter is concentrated near the white-dwarf surface.

\section{Fitting a data set: Application to V834 Cen}
\label{v834}

To explore our model and test the fitting procedure of CYCLOPS, we have applied it to multiwavelength observations of a polar (V834 Cen) found in the literature. This aims to illustrate the potential of our tool and also to trace the path to future improvements.

V834 Centauri is an object already recognized as having extended emitting regions (e.g., \citealt{cro89,pot04}). The reader is referred to these works (and references therein) to a description of the system. From the many published Stokes curves of V834 Cen, we have chosen those presented in \citet{bai83}. They represent the magnitudes and circular polarization of V834 Cen in optical and near-infrared bands and are ideal to test our fitting procedure using two data set in different harmonic regions. Specifically, we have used the V and J band curves converted to flux and polarized flux. We have used frequencies corresponding to the central wavelength of the bands.

\begin{figure}
\includegraphics[width=84mm]{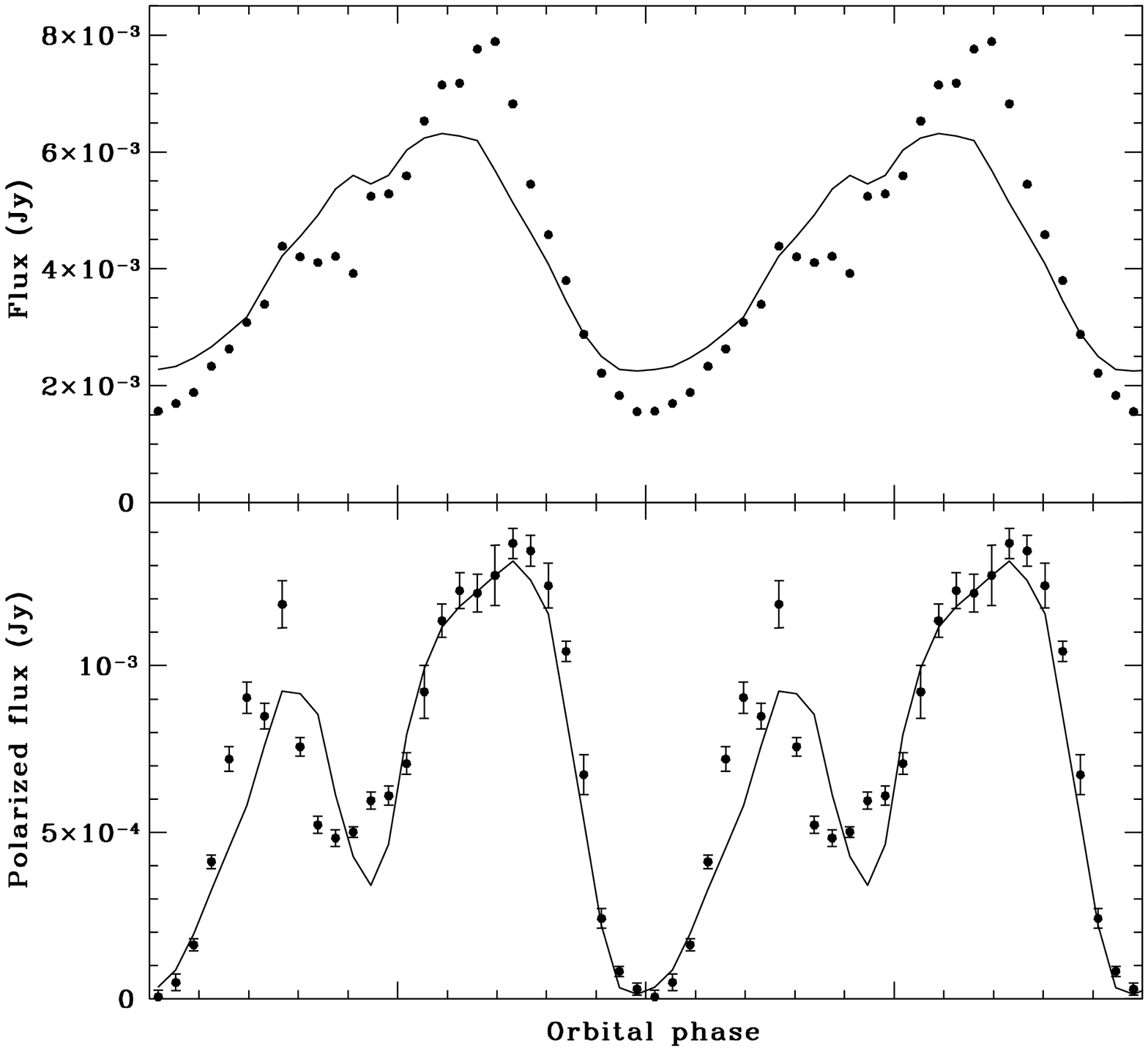} 
\includegraphics[width=84mm]{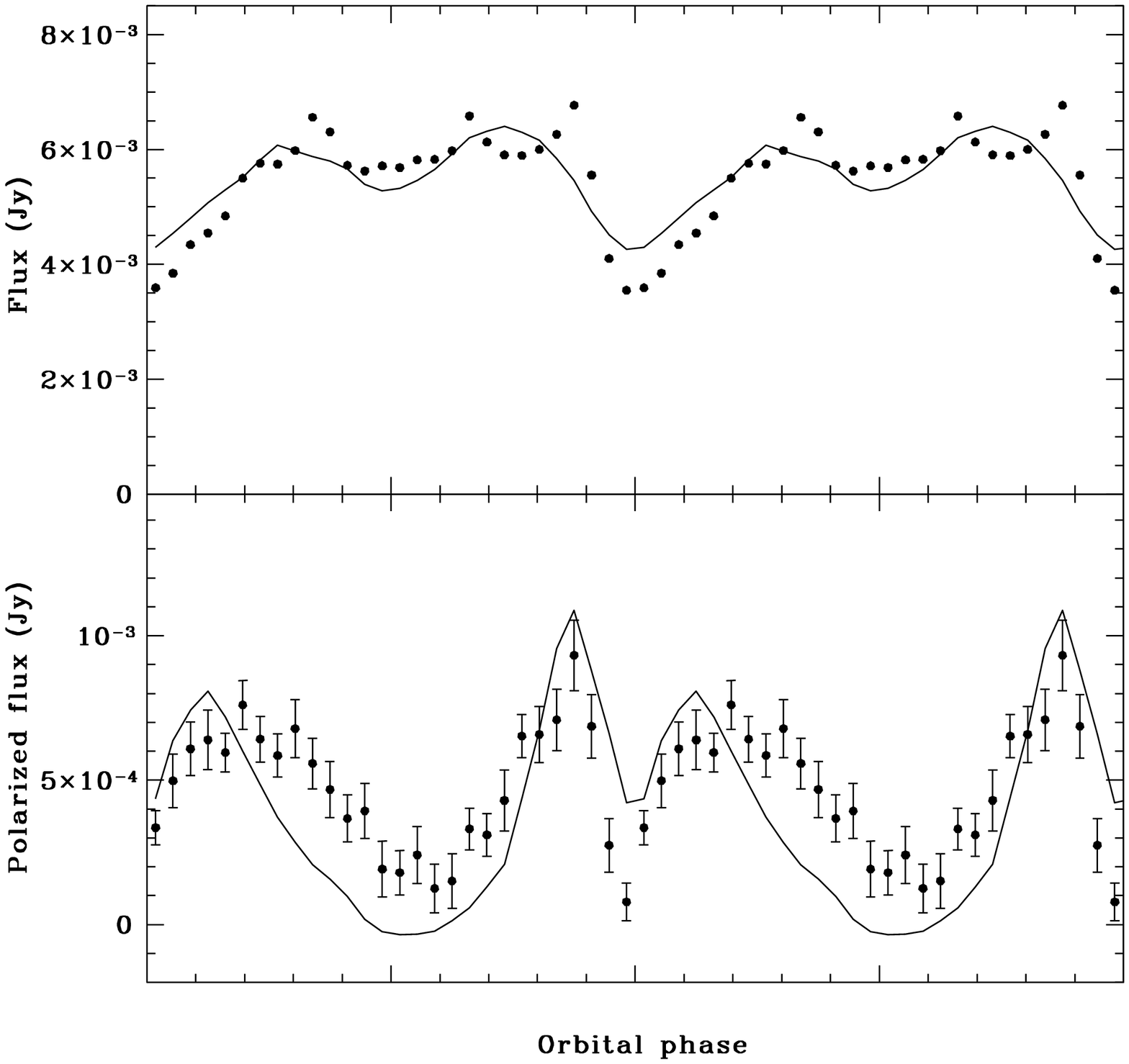} 
$ $\vspace{3cm}
\caption{Best fit to the data of V834 Cen from \citet{bai83}. The V band is on the left panel and the J band, on the right. See Table \ref{tab_param} for the model parameters.}
\label{fig_v834}
\end{figure}

\begin{figure}
$ $\vspace{3cm}
\includegraphics[width=84mm]{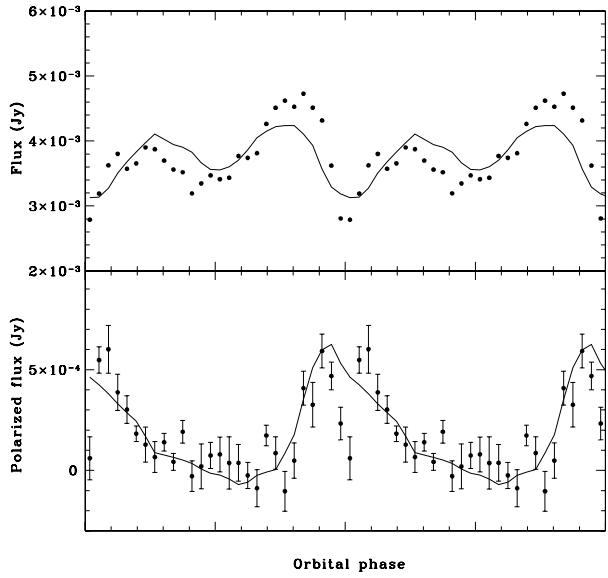} 
\caption{Resulting curve in H band using the fitted parameters of V834 Cen superposed to the data from \citet{bai83}.}
\label{fig_v834_h}
\end{figure}

We could not find a good simultaneous fit to both bands using constant values of density and temperature in tangential direction. So we adopted a lateral variation as explained in Section \ref{sec_defining_region}. Figure \ref{fig_v834} presents the model with the better value of $\chi^2$. It was found from a search with no restriction in the parameter space. It produces a huge number of possibilities, but we think it is a better approach for testing the search algorithm. The parameter of this model is presented in Table \ref{tab_param}. The gross features are reproduced, including the relative fluxes between the bands. We suggest that a different matter distribution in the emitting region may produce a finer agreement with the data. The above model also reproduces the H data of V834 Cen \citep{bai83}, even if not included in the fit procedure (Figure \ref{fig_v834_h}).

\begin{table}
\caption{Best fitting to V834 Cen (see text for details).}
\label{tab_param}
\begin{tabular}{lc}
\hline
Parameter & Value\\
\hline
$i$ (\degr) & 44 \\
$\beta$ (\degr) & 40 \\
$\Delta R$ & 0.132  \\
$\Delta long$ (\degr) & 47 \\
%, one-half of the azimuth extension of the threading region: 49.1\degr;
$h$ & 0.19 \\
%, the length of the emitting region in the WD radial direction: 0.08;
$f_l$ & 0.254\\
$R_{th}$ ($R_{WD}$) & 6.1 \\
%the colatitude of the emitting region: 32.2\degr. This results in a threading region centred at 24 $R_{WD}$;
Limits of the threading region in radius ($R_{WD}$) & 5.32 - 6.9 \\
%, one-half of the fractional radial extension of the threading region: 0.36. It corresponds to a threading region extending from 15.3 to 32.5 $R_{WD}$. ;
$B_p$ (MG) & 34.4 \\
Range of $B$ in the emitting region (MG): & 18.8 - 33.2 \\
% the magnetic field intensity in the {\it anchor point}: 3.8 MG ;
$B_{lat}$ (\degr) & 46  \\
%, the latitude of the magnetic axis: 56\degr;
$B_{long}$ (\degr) & 327 \\
%, the longitude of the magnetic axis: 21\degr;
$N_e$, maximum electronic density (cm$^{-3}$) & $8.1\ 10^{11}$ \\
$T$, maximum electronic temperature (keV) and average temperature & 171.3 / 20 \\
Unpolarised component (Jy) & 0.0022 (V) - 0.0038 (J) - 0.0026 (H)  \\ 
$\Delta$phase & -0.069  \\
Attenuation by the upper column & 0 \\
\hline
\end{tabular}

%$^*$ This parameter is not fitted.
\end{table}

The parameters relative to the geometry of V834 Cen are consistent with previous fittings of the system. For instance, the previous inclinations found in the literature are in the range 40 -- 60 \degr \citep{cro89,sch93,fer90}. Extended emitting region was previous found \citep{cro89,fer90,pot04}. This is naturally explained by an extended threading region. Previous estimates to its radius can be found in \citet[from 5.9 to 37.5 $R_{WD}$]{mau02}, using ultraviolet photometry and spectroscopy, and in \citet{cro89}, who also found values of order of 30 $R_{WD}$ using polarimetric data. We can find a number of estimates to the temperature using high-energy data: they concentrate in the range 10 - 20 keV \citep[e.g.]{bar06}. Our average temperature is also consistent to those values. The magnetic field estimate from Zeeman line splitting in white dwarf photosphere is 23 MG \citet{fer92}. 

The number of possible models, even for narrow bands of all free parameters, is huge. Thus, the search for the best model is limited by the available time and processing power. The genetic algorithm has proven to be a robust and efficient method. Even so, a very large database of models was generated. It is hard to quantify how much has been explored from the universe of possibilities, but the solution we found is an improvement in the sense it simultaneously explain three wavebands.

\section{Conclusions}

We present a new code, CYCLOPS, to study the optical emission from AM Herculis systems. It calculates the cyclotron emission from a 3D region on or near the white-dwarf surface. The emitting region is presented by a 3D grid with physical properties spatially variable but physically connected by a function. The adopted solution by steps for the radiative transfer is adequate to represent optically thick columns. The code also incorporates a procedure to fit an observational data set. An application to previous published multiband V834 Cen data is shown. A model with an extended region with electronic temperature and density varying with height and in the tangential direction is able to account for the data. 

\section*{ACKNOWLEDGEMENTS}

We acknowledge the referees by their suggestions which were valuable in improving the final version of the manuscript. This work was partially supported by Fapesp (CVR: Proc. 2001/12589-1). This research has made use of the NASA's Astrophysics Data System Service.

\clearpage

\clearpage

% \clearpage

\clearpage

\clearpage

\clearpage

\clearpage

\appendix

\section{A new expression for the solution of cyclotron radiative transfer}
\label{apendice}

The general solution of the transfer equation for the Stokes parameters in a magnetoactive plasma is given by \citet{pac75}. Unfortunately, there are misprints in equations 7 and 13. Therefore, we consider useful to show them again. The correct form of equation 7 of \citet{pac75} is:

\begin{equation}
 \begin{array}{lcl}
  I_H & = & \sum\limits_{i=1}^4 K_i \exp ({\alpha_i s}) , \\
  Q_H & = & - {\displaystyle \frac{1}{r}} \sum\limits_{i=1}^4 \alpha_i (v a_{i+2} + h) K_i \exp ({\alpha_i s}) , \\
  U_H & = & - \sum\limits_{i=1}^4 a_{i+2} K_i \exp ({\alpha_i s}) , \\
  V_H & = & {\displaystyle \frac{1}{r}} \sum\limits_{i=1}^4 \alpha_i (q a_{i+2} - f) K_i \exp ({\alpha_i s}) .
 \end{array}
\end{equation}

The constants $K_i$ (\citealt{pac75}, eq. 13) should be written as:

\begin{equation}
 \begin{array}{lcl}
  K_{1,2} & = & {\displaystyle \frac{p}{4[(m-n)^2+r^2]^{1/2}} \left[ a_i (I_o - I_p) - \frac{q a_i - f}{\alpha_i} (Q_o-Q_p) + (U_o - U_p) - \frac{v a_i + h}{\alpha_i} (V_o - V_p) \right]} , \\
  K_{3,4} & = & - K_{1,2} .
 \end{array}
\end{equation}

We refer the reader to the original paper of \citet{pac75} for the definitions of the symbols used. 

The above expressions were checked as a solution of the differential equation. They also provide the solution of \citet{meg82} for the case with no input radiation ($I_o=Q_o=U_o=V_o=0$).

\vspace{1.5cm}
Now, we present an alternative representation for the complete solution of \citet{pac75} expressed in terms of trigonometric and hyperbolic functions. 

\begin{equation}
\label{sin_cos}
\begin{array}{lcl}
  I(s) & = & 2 { \left[ K_1^A \cosh{(\lambda s)} + K_1^B \sinh{(\lambda s)} + K_3^A \cos{(\mu s)} + K_3^B \sin{(\mu s)} \right] }e^{-\kappa s} + I_p , \\
\\
  Q(s) & = & -2 \left\{ \displaystyle \frac{q a_1 -f}{\lambda a_1} \left[ K_1^A \sinh (\lambda s) + K_1^B \cosh (\lambda s)\right] + \frac{\mu}{r} \left( v a_1 +h \right) \left[ - K_3^A \sin (\mu s) + K_3^B \cos (\mu s) \right] \right\} e^{-\kappa s} + Q_p  , \\
\\
  U(s) & = & -2 \left\{ a_3 \left[ K_1^A \cosh{(\lambda s)} + K_1^B \sinh{(\lambda s)} \right] + a_1 \left[ K_3^A \cos{(\mu s)} + K_3^B \sin{(\mu s)} \right] \right\} e^{-\kappa s} + U_p , \\
\\
  V(s) & = & 2 \left\{ \displaystyle -\frac{h+v a_1}{\lambda a_1} \left[ K_1^A \sinh (\lambda s) + K_1^B \cosh (\lambda s)\right] + \frac{\mu}{r} \left( q a_1 -f \right) \left[ - K_3^A \sin (\mu s) + K_3^B \cos (\mu s) \right] \right\} e^{-\kappa s} + V_p  ,
\end{array}
\end{equation}

\noindent where

\begin{equation}
\label{eq_K}
\begin{array}{lcl}
K_1^A & = & {\displaystyle \frac{p}{4 R} \left[ a_1 (I_o - I_p) + (U_o - U_p) \right]} , \\
\\
K_3^A & = & {\displaystyle -\frac{p}{4 R} \left[ \frac{1}{a_1} (I_o - I_p) + (U_o - U_p) \right]} , \\
\\
K_1^B & = & - {\displaystyle \frac{p}{4 R} \left[ \frac{q a_1 - f}{\lambda} (Q_o-Q_p) + \frac{v a_1 + h}{\lambda} (V_o - V_p) \right]} , \\
\\
K_3^B & = & - {\displaystyle \frac{p}{4 R} \frac{\mu}{r} \left[ \left( v+ \frac{h}{a_1} \right) (Q_o-Q_p) - \left( q - \frac{f}{a_1} \right) (V_o - V_p) \right]} .
\end{array}
\label{solution}
\end{equation}

In the above expressions, we have used the following identities:

\[
\begin{array}{lcl}
\alpha_1 & = & \lambda , \\
\alpha_3 & = & i \mu, \\
R & = & [(m-n)^2+r^2]^{1/2},
\end{array}
\]

\noindent where $\lambda$, $\mu$, and $R$ are definitions presented in \citet{meg82}.

In the thermal regime, the particular solution is the Planck function: $[I_p,Q_p,U_p,V_p] = [B_w,0,0,0]$. 

Equations \ref{sin_cos} and \ref{eq_K} are easily reduced to the solution presented in \citet{meg82} for the case with no incident radiation, $[I_o,Q_o,U_o,V_o] = 0$.

\end{document}